\documentstyle[12pt]{article}

\makeatletter

 \@addtoreset{equation}{section}
\makeatother

\begin{document}
\topmargin 0pt
\oddsidemargin 5mm

\newcommand {\beq}{\begin{eqnarray}}
\newcommand {\eeq}{\end{eqnarray}}
\newcommand {\non}{\nonumber\\}
\newcommand {\eq}[1]{\label {eq.#1}}
\newcommand {\defeq}{\stackrel{\rm def}{=}}
\newcommand {\gto}{\stackrel{g}{\to}}
\newcommand {\hto}{\stackrel{h}{\to}}
\newcommand {\1}[1]{\frac{1}{#1}}
\newcommand {\2}[1]{\frac{i}{#1}}
\newcommand {\th}{\theta}
\newcommand {\thb}{\bar{\theta}}
\newcommand {\ps}{\psi}
\newcommand {\psb}{\bar{\psi}}
\newcommand {\ph}{\varphi}
\newcommand {\phs}[1]{\varphi^{*#1}}
\newcommand {\sig}{\sigma}
\newcommand {\sigb}{\bar{\sigma}}
\newcommand {\Ph}{\Phi}
\newcommand {\Phd}{\Phi^{\dagger}}
\newcommand {\Sig}{\Sigma}
\newcommand {\Phm}{{\mit\Phi}}
\newcommand {\eps}{\varepsilon}
\newcommand {\del}{\partial}
\newcommand {\dagg}{^{\dagger}}
\newcommand {\pri}{^{\prime}}
\newcommand {\prip}{^{\prime\prime}}
\newcommand {\pripp}{^{\prime\prime\prime}}
\newcommand {\prippp}{^{\prime\prime\prime\prime}}
\newcommand {\pripppp}{^{\prime\prime\prime\prime\prime}}
\newcommand {\delb}{\bar{\partial}}
\newcommand {\zb}{\bar{z}}
\newcommand {\mub}{\bar{\mu}}
\newcommand {\nub}{\bar{\nu}}
\newcommand {\lam}{\lambda}
\newcommand {\lamb}{\bar{\lambda}}
\newcommand {\kap}{\kappa}
\newcommand {\kapb}{\bar{\kappa}}
\newcommand {\xib}{\bar{\xi}}
\newcommand {\ep}{\epsilon}
\newcommand {\epb}{\bar{\epsilon}}
\newcommand {\Ga}{\Gamma}
\newcommand {\rhob}{\bar{\rho}}
\newcommand {\etab}{\bar{\eta}}
\newcommand {\chib}{\bar{\chi}}
\newcommand {\tht}{\tilde{\th}}
\newcommand {\zbasis}[1]{\del/\del z^{#1}}
\newcommand {\zbbasis}[1]{\del/\del \bar{z}^{#1}}
\newcommand {\vecv}{\vec{v}^{\, \prime}}
\newcommand {\vecvd}{\vec{v}^{\, \prime \dagger}}
\newcommand {\vecvs}{\vec{v}^{\, \prime *}}
\newcommand {\alpht}{\tilde{\alpha}}
\newcommand {\xipd}{\xi^{\prime\dagger}}
\newcommand {\pris}{^{\prime *}}
\newcommand {\prid}{^{\prime \dagger}}
\newcommand {\Jto}{\stackrel{J}{\to}}
\newcommand {\vprid}{v^{\prime 2}}
\newcommand {\vpriq}{v^{\prime 4}}
\newcommand {\vt}{\tilde{v}}
\newcommand {\vecvt}{\vec{\tilde{v}}}
\newcommand {\vecpht}{\vec{\tilde{\phi}}}
\newcommand {\pht}{\tilde{\phi}}
\newcommand {\goto}{\stackrel{g_0}{\to}}
\newcommand {\tr}{{\rm tr}\,}
\newcommand {\GC}{G^{\bf C}}
\newcommand {\HC}{H^{\bf C}}
\newcommand{\vs}[1]{\vspace{#1 mm}}
\newcommand{\hs}[1]{\hspace{#1 mm}}
\newcommand{\al}{\alpha}
\newcommand{\be}{\beta}
\newcommand{\Lam}{\Lambda}

\setcounter{page}{0}

\begin{titlepage}

\begin{flushright}
OU-HET 386\\
TIT/HEP-464\\
{\tt hep-th/0106183}\\
June 2001
\end{flushright}
\bigskip

\begin{center}
{\LARGE\bf
Supersymmetric Extension of 
the Non-Abelian Scalar-Tensor Duality
}
\vs{10}

\bigskip
{\renewcommand{\thefootnote}{\fnsymbol{footnote}}
\large\bf 
Ko Furuta$^1$\footnote{
E-mail: furuta@phys.chuo-u.ac.jp}, 
Takeo Inami$^1$, 
Hiroaki Nakajima$^1$\footnote{
E-mail: nakajima@phys.chuo-u.ac.jp}\\
and Muneto Nitta$^2$\footnote{
E-mail: nitta@het.phys.sci.osaka-u.ac.jp}
}

\setcounter{footnote}{0}
\bigskip

{\small \it
$^1$ Department of Physics, Chuo University, 
Tokyo 112-8551, Japan\\
$^2$ Department of Physics,
Graduate School of Science, Osaka University,\\
Toyonaka
 560-0043, Japan\\
}

\end{center}
\bigskip

\begin{abstract}
The field theory dual to the Freedman-Townsend model 
of a non-Abelian anti-symmetric tensor field 
is a nonlinear sigma model on the group manifold $G$. 
This can be extended to the duality between  
the Freedman-Townsend model coupled to Yang-Mills fields 
and a nonlinear sigma model on a coset space $G/H$. 
We present the supersymmetric extension of this duality 
and find that the target space of this 
nonlinear sigma model is a complex coset space, $\GC/\HC$.
 
\end{abstract}

\end{titlepage}

\section{Introduction}
Electric-magnetic duality plays a crucial role 
in studies of non-perturbative dynamics of 
four-dimensional supersymmetric gauge theories~\cite{SUSY-QCD}. 
In four dimensions, 
there exists another duality between the scalar field 
and the anti-symmetric tensor (AST) field, 
called the scalar-tensor duality~\cite{KR}. 
The AST field 
$B(x) = \1{2} B_{\mu\nu}(x)dx^{\mu} \wedge dx^{\nu}$ 
is an important ingredient 
in supergravity and superstrings~\cite{CJ,string}. 
The massless theory of the AST field possesses 
AST gauge invariance:   
$B(x) \to B(x) + d \xi(x)$, 
with $\xi (x)$ being a one-form gauge parameter.   
In this case, the dual field theory is a free scalar field theory.

The non-Abelian generalization of the theory of the AST field 
was made by Freedman and Townsend~\cite{FT}, 
and earlier by Ogievetsky and Polubarinov~\cite{OP}.
It is called the `Freedman-Townsend (FT) model'. 
In this theory, the AST field
$B_{\mu\nu} = B^i{}_{\mu\nu} T_i$ transforms  
as the adjoint representation of the group $G$, 
where $T_i$ are generators of $G$.  
This theory possesses a non-Abelian generalization of 
the AST gauge invariance~\cite{FT,SOS}.  
In this case, the dual scalar field theory is  
a nonlinear sigma model (NLSM) on the group manifold $G$,  
called the principal chiral model. 
Nonlinear sigma model in four dimensions  
have attracted interest as low-energy effective 
field theories of QCD. 
The NLSM on the coset spaces $G/H$ 
naturally appear when the global symmetry is spontaneously
broken down to its subgroup $H$~\cite{CCWZ}. 
The scalar-tensor duality can be extended 
to a duality between 
AST gauge theories 
coupled to the Yang-Mills (YM) gauge field 
and NLSM on coset spaces $G/H$~\cite{FrT}. 

Supersymmetry plays roles in realizing 
electric-magnetic duality and other types of dualities 
at the quantum level in gauge theories and 
string theories~\cite{SUSY-QCD,string}.
In this paper, we investigate the supersymmetric extension 
of the scalar-tensor duality.  
The Abelian scalar-tensor duality in 
supersymmetric field theories 
has been studied extensively~\cite{Si,LR,HLR}. 
The non-Abelian generalization of scalar-tensor duality 
in supersymmetric field theories was first reported 
in Ref.~\cite{CLL}. 
The dual field theories are 
the supersymmetric NLSM 
on the complex extension of the group manifold $G$, $\GC$. 
This is related to the fact that 
the target spaces of the supersymmetric NLSM 
must be K\"ahler~\cite{Zu}. 
We extended the previous study of scalar-tensor duality in 
supersymmetric theories from a group manifold $\GC$ to the 
coset spaces $\GC/\HC$

This paper is organized as follows. 
In \S2, we recapitulate the 
non-Abelian scalar-tensor duality in bosonic theories. 
In \S3, we recall the scalar-tensor duality 
in supersymmetric field theories in the case of 
the complex extension of the group manifold as the target space of 
the NLSM~\cite{CLL}.  
The relation between the (quasi-) Nambu-Goldstone bosons 
in the NLSM and the boson fields in the AST gauge theory 
is clarified.  
In \S4 we construct the supersymmetric extension of 
the duality of the non-Abelian AST field theory 
and the NLSM on a coset space on the basis of 
the results in \S2 and \S3. 
Section 5 is devoted to discussion.

\section{Non-Abelian scalar-tensor duality in bosonic theories}
We begin by recapitulating the non-Abelian scalar-tensor 
duality~\cite{FT} and its $G/H$ generalization~\cite{FrT}. 
We consider the non-Abelian AST field 
$B_{\mu\nu}$ with the group $G$.
Our notation for its Lie algebra ${\cal G}$ is 
\beq
 && T_i \in {\cal G},\hs{10} i =1,\cdots,\dim G ,\non
 && [T_i,T_j] = i f_{ij}{}^k T_k, \hs{10} 
    \tr (T_i T_j) = c \delta_{ij}, \label{Lie_alg.}
\eeq
where the coefficients $f_{ij}{}^k$ are the structure constants, 
and $c$ is a positive constant. 
We use matrix notation for 
the AST field and 
the auxiliary vector field: 
\beq
 B_{\mu\nu}(x) \equiv B^i_{\mu\nu}(x)T_i, \hs{10}
 A_{\mu}(x) \equiv A^i_{\mu}(x)T_i. \label{matrix}
\eeq
The mass dimensions of $B_{\mu\nu}$ and $A_{\mu}$ are $1$ and $2$, 
respectively. 
The Lagrangian of the FT model 
in first-order form is given by
\beq
 {\cal L} = - \frac{1}{8c}{\rm tr} 
 \left(\epsilon^{\mu\nu\rho\sigma}B_{\mu\nu}
   F_{\rho\sigma} + A_{\mu}A^{\mu}\right),\label{1-order}
\eeq
where
$F_{\rho\sigma}\equiv\partial_\rho A_\sigma 
-\partial_\sigma A_\rho - i \lambda
\left[ A_\rho,A_\sigma\right]$, and 
$\lam$ is the coupling constant, with dimension $-1$.
This Lagrangian is invariant up to a total derivative under 
the AST gauge transformation, defined by 
\beq
 \delta B_{\mu\nu} 
   = D_{\mu}(A) \xi_{\nu} - D_{\nu}(A) \xi_{\mu}, \hs{10}
 \delta A_{\mu} = 0, \label{AST-tr.1}
\eeq
where $\xi_{\mu} (x) \equiv \xi^i_{\mu}(x)T_i$ is 
a vector field gauge parameter, 
and we have introduced the covariant derivative 
$D_{\mu}(A) \xi_{\nu} = \partial_{\mu}\xi_{\nu} 
- i \lambda [A_{\mu}, \xi_{\nu}]$. 
In addition, the Lagrangian (\ref{1-order}) 
has the {\it global} symmetry of the group $G$ 
under the transformation
\beq
 B_{\mu\nu}\to B'_{\mu\nu}=g^{-1}B_{\mu\nu}g , \hs{10}
 A_{\mu}\to A'_{\mu}= g^{-1}A_{\mu}g, \label{Gglobe}
\eeq
with $g\in G$. 

The second-order Lagrangian for $B^i_{\mu\nu}$ is 
obtained by eliminating $A^i_\mu$ (see Ref.~\cite{FT}). 
We obtain the dual formulation 
of the AST gauge theory by 
eliminating $B^i_{\mu\nu}$ instead. 
This gives the flatness condition for $A_\mu$, 
$F_{\mu\nu}=0$. The solution of this equation is given by
\beq
  A_{\mu} = {i \over{\lambda}} 
            U^{-1}\partial_{\mu}U,\label{solution}
\eeq
where $U(x) \in G$. 
Substituting (\ref{solution}) back into 
the Lagrangian (\ref{1-order}), we obtain
\beq
  {\cal L}= \frac{1}{8c\lambda^2} 
  {\rm tr} \left(U^{-1}\partial_\mu U \right)^2.\label{NLSM}
\eeq
This is the NLSM on the group manifold $G$, 
called the principal chiral model. 

The generalization of the scalar-tensor duality to the case of 
a coset space $G/H$ as the target space of the NLSM can be 
made by introducing the YM field for 
the subgroup $H$~\cite{FrT},
\beq
  v_\mu\equiv v_\mu^a H_a \in{\cal H}.
\eeq
Here we have decomposed the Lie algebra ${\cal G}$ into 
the subalgebra ${\cal H}$ and the coset generators as 
\beq
    T_i \in {\cal G}, \hs{10}
    H_a \in {\cal H}, \hs{10} 
    X_I \in {\cal G} -{\cal H} , \hs{10}
  \tr (H_a X_I) = 0. \label{generator}
\eeq
The AST field and the auxiliary vector field 
are defined in the same way as in (\ref{matrix}). 
The YM gauge transformations of the group $H$ now read
\begin{eqnarray}
 &&
 v_\mu\,\,\,\to\,\,\, v'_\mu = h^{-1}(x) 
   \left(v_\mu + {i\over e}\partial_\mu\right) h(x),\non
 && B_{\mu\nu} \to B'_{\mu\nu} = h^{-1}(x)B_{\mu\nu}h(x), \hs{10}
  A_\mu\,\,\,\to\,\,\,A'_\mu = h^{-1}(x)A_\mu h(x), 
 \label{bosonic_YM}
\end{eqnarray}
with $h(x)\in H$. 
Here $e$ is the YM coupling constant with dimension 0. 
The first-order Lagrangian invariant under 
this gauge transformation is given by 
\beq
 {\cal L} = - \frac{1}{8c}\epsilon^{\mu\nu\rho\sigma}
 {\rm tr}\left[ B_{\mu\nu}
  F_{\rho\sigma}\left(A+\frac{e}{\lambda}v\right)\right]
  - \frac{1}{8c}{\rm tr}(A_\mu A^\mu), 
  \label{coset1'}
\eeq
where $F_{\mu\nu} \left( A+ \frac{e}{\lambda}v \right)\equiv 
F_{\mu\nu} + \frac{e}{\lambda}
\left(\partial_\mu v_\nu-\partial_\nu v_\mu
-i e\left[v_\mu,v_\nu\right]\right) 
-i e\left(\left[ A_\mu, v_\nu \right]
+\left[v_\mu, A_\nu \right]
\right)$.

The AST gauge transformation is modified from (\ref{AST-tr.1}) 
after taking account of the coupling of $B^i_{\mu\nu}$ 
with the YM field $v^a_{\mu}$. 
One only has to replace the covariant derivative 
$D_{\mu}(A)$ by $D_{\mu}(A + \frac{e}{\lambda} v)$: 
\begin{eqnarray}
 && \delta B_{\mu\nu} = 
   D_{\mu}\left(A + \frac{e}{\lambda} v\right) \xi_{\nu} 
 - D_{\nu}\left(A + \frac{e}{\lambda} v\right) \xi_{\mu}, \non 
 && \delta A_\mu = 0, \hs{10}
    \delta v_\mu = 0. \label{AST-GT2}
\end{eqnarray}
The case in which $H$ is $G$ itself is considered in Ref.~\cite{FT}. 
The model used in that work is different from ours in the sense that 
it contains the kinetic term of the YM field. 

Note that the existence of the gauge field $v_\mu$ in 
the Lagrangian (\ref{coset1'}) 
explicitly breaks the original $G$ symmetry of the Lagrangian (\ref{1-order}), 
since $v'_\mu=g^{-1} v_\mu g$ is not necessarily an element of 
${\cal H}$.

Eliminating $A_\mu$ from the Lagrangian (\ref{coset1'}),
we obtain the second-order Lagrangian for $B_{\mu\nu}$ coupled to 
the YM field $v_\mu$, 
\beq
 {\cal L} = 
 - \1{8} \left(\tilde G^{\mu i} \tilde K^{ij}_{\mu\nu} \tilde G^{\nu j} 
 + {e \over \lam} \epsilon^{\mu\nu\rho\sig} 
   B^i_{\mu\nu} v_{\rho\sig}^i \right), \label{sec}
\eeq
where we have defined the field strength 
$\tilde G^{\mu i} \equiv 
  \epsilon^{\mu\nu\rho\sig} D_{\nu} B^i_{\rho\sig} 
  \equiv \epsilon^{\mu\nu\rho\sig} 
 (\del_{\nu}B^i_{\rho\sig} 
  + e f^{ijk} v^j_{\nu} B^k_{\rho\sig})$,
and $v^i_{\mu\nu}$ is the YM field strength. 
Here, $\tilde K_{\mu\nu}^{ij}$ is related to  
$K^{\mu\nu ij} \equiv 
  g^{\mu\nu}\delta^{ij} 
  - \lam f^{ijk}\epsilon^{\mu\nu\rho\sig}B^k_{\rho\sig}$ 
through 
$\tilde K_{\mu\rho}^{ik} K^{\rho\nu kj} 
  = \delta^{\nu}_{\mu} \delta^{ij}$.

We now proceed to the derivation of the dual scalar field theory 
equivalent to the AST field theory (\ref{sec}). 
Elimination of $B^i_{\mu\nu}$ in 
the Lagrangian (\ref{coset1'}) yields the constraint
\beq
 F_{\mu\nu}\left(A+\frac{e}{\lambda}v\right)=0.\label{flatness}
\eeq
Its solution is  
\beq
 A_{\mu} + \frac{e}{\lambda}v_\mu 
 ={i\over \lambda} U(x)^{-1}\partial_\mu U(x), 
  \quad U(x)\in G.  \label{solution2}
\eeq
Substituting this expression back into Eq.~(\ref{coset1'}), 
we obtain
\beq
 {\cal L} = \frac{1}{8c\lambda^2}{\rm tr}(U^{-1}D_\mu U)^2,
 \label{dualcoset2}
\eeq
where we have introduced the covariant derivative, 
$D_\mu U\equiv \partial_\mu U + ie Uv_\mu$.
The Lagrangian (\ref{dualcoset2}) possesses the 
{\it global} $G_L \times$ {\it local} $H_R$ symmetry 
[where the subscript $L$ ($R$) refers to left (right) action], given by
\beq
 U(x)\to gU(x),\quad v_\mu(x) \to v_\mu(x),\label{globalG}
\eeq
with $g\in G$, and
\beq
 U(x) \to U(x)h(x),\quad 
 v_\mu \to h^{-1}(x)\left(v_\mu + {i\over e}\partial_\mu\right)h(x),
 \label{localH}
\eeq
with $h(x)\in H$.
The global $G$ transformation (\ref{globalG}) is 
hidden in the AST gauge theory. 
Eliminating the auxiliary field $v_\mu$ by using the equation of 
motion, we obtain the usual form 
of the NLSM on $G/H$~\cite{CCWZ,BKY}.

There is an alternative way of writing the first-order Lagrangian; 
we replace the auxiliary field $A_{\mu}$ as 
\beq
 A_{\mu} \to A_{\mu} - {e\over \lam} v_{\mu}, \label{redef}
\eeq
by field redefinition, and 
the Lagrangian (\ref{coset1'}) then becomes 
\beq
 {\cal L} = - \frac{1}{8c}\epsilon^{\mu\nu\rho\sigma}
 {\rm tr}\left(B_{\mu\nu}
  F_{\rho\sigma} \right)
  - \frac{1}{8c}{\rm tr} 
  \left[\left(A_{\mu} - {e\over \lam} v_{\mu}\right) 
        \left(A^{\mu} - {e\over \lam} v^{\mu}\right) \right]. 
  \label{coset1''}
\eeq
This expression of the Lagrangian is obtained in Ref.~\cite{FrT}. 
Note that it is invariant under 
the AST gauge transformation (\ref{AST-tr.1}) with 
$\delta v_{\mu} = 0$, instead of (\ref{AST-GT2}). 
The YM gauge transformation 
of $A_{\mu}$ takes the usual form: 
$A_\mu \to A'_\mu 
 = h^{-1}(x)\left(A_\mu + {i\over \lam}\partial_\mu\right)h(x)$.

\section{Supersymmetric extension of the non-Abelian 
scalar-tensor duality for group manifolds}
In this section, after we recapitulate the non-Abelian scalar-tensor duality 
for the group manifold in 
supersymmetric field theories~\cite{CLL}, 
we discuss the correspondence of fields in the two theories.

In supersymmetric field theories, 
the AST field $B_{\mu\nu}(x)$ 
is a component of the (anti-)chiral spinor superfields 
$B_{\al}(x,\th,\thb)$ [$\bar B_{\dot\al}(x,\th,\thb)$], 
satisfying the supersymmetric constraints~\cite{Si}
\beq
 \bar D_{\dot\al} B_{\be}(x,\th,\thb) = 0, \hs{10} 
 D_{\al} \bar B_{\dot\be}(x,\th,\thb) = 0.
\eeq
These superfields can be expanded in terms of component fields as 
\beq
&& \hs{-5} 
 B^{\al}(y,\th) 
 = \psi^{\al}(y) + \1{2} \th^{\al}(C(y) + i D(y)) 
  + \1{2} (\sig^{\mu\nu})^{\al\be} \th_{\be} B_{\mu\nu}(y)
  + \th\th\lam^{\al}(y), \non
&& \hs{-5} 
 \bar B_{\dot\al}(y\dagg,\bar\th) 
 = \psb_{\dot\al}(y\dagg) 
  + \1{2} \thb_{\dot\al}(C(y\dagg) - i D(y\dagg)) 
  +\1{2} (\bar\sig^{\mu\nu})_{\dot\al\dot\be} \thb^{\dot\be} 
    B_{\mu\nu}(y\dagg)
  + \thb\thb\bar\lam_{\dot\al}(y\dagg), \hs{10}
\eeq
where $y^{\mu} = x^{\mu} + i\th \sig^{\mu} \thb$ 
and $\bar D_{\dot\al} = \del/\del \bar\th^{\dot\al}$ 
(or $y^{\mu\dagger} = x^{\mu} - i\th \sig^{\mu} \thb$ 
and $D_{\al} = - \del/\del\th^{\al}$), 
and 
$(\sig^{\mu\nu})^{\al}{}_{\beta} 
= \1{4} (\sig^{\mu}\sigb^{\nu} 
- \sig^{\nu}\sigb^{\mu})^{\al}{}_{\beta}$ (see Ref.~\cite{WB}). 
The mass dimension of $B_{\al}$ is $\1{2}$.\footnote{
Here we list the dimensions of the (super)fields 
introduced in this and the following sections:  
$[B_{\al}]=\1{2}$, $[B_{\mu\nu}]=1$, 
$[A]=1$, $[A_{\mu}]=2$, $[W_{\al}]={5 \over 2}$, 
$[V]=0$ and $[v_{\mu}]=1$. 
Here, $V$ is the YM vector superfield 
introduced in the next section. 
The dimensions of the coupling constants are 
$[\lam]=-1$ and $[e]=0$, as in the bosonic case.
}

We consider the non-Abelian AST field 
$B_{\mu\nu}$ with the group $G$ [see (\ref{Lie_alg.})]. 
This field $B_{\mu\nu}$ is a component of ${\cal G}$-valued 
(anti-)chiral spinor superfields 
$B_{\al}(x,\th,\thb)=B^i_{\al}(x,\th,\thb) T_i$ 
[$\bar B_{\dot\al}(x,\th,\thb) = \bar B^i_{\dot\al}(x,\th,\thb) T_i$]. 

To construct the supersymmetric extension of the FT model, 
we introduce a ${\cal G}$-valued auxiliary vector superfield 
$A(x,\th,\thb) = A^i(x,\th,\thb) T_i$, 
satisfying the constraint ${A^i}\dagg = A^i$. 
Its field strengths are (anti-)chiral spinor superfields,  
\beq
 W_{\al} = - \1{4 \lam} \bar D \bar D 
             (e^{-\lam A}D_{\al}e^{\lam A}) ,\hs{10}
 \bar W_{\dot\al} 
  = \1{4 \lam} D D (e^{\lam A} \bar D_{\dot\al}e^{-\lam A}). 
 \label{Auxiliary-fs}
\eeq
The first-order Lagrangian can be written as~\cite{LR,CLL}\footnote{
See the discussion in \S5 concerning variants of the Lagrangian 
(\ref{Lag.}).
}  
\beq
 {\cal L} = - \1{2c} \left[\int d^2\th \; \tr (W^{\al} B_{\al}) 
 + \int d^2 \thb \; \tr (\bar W_{\dot\al} \bar B^{\dot\al}) \right]
 + \1{4c} \int d^4 \th \; \tr A^2. \label{Lag.}
\eeq
This Lagrangian is verified by constructing 
the supersymmetric AST gauge transformation 
which leaves it invariant. 
To this end we first define the covariant spinor derivative 
${\cal D}_{\al} = D_{\al} 
+ [e^{-\lam A} D_{\al}e^{\lam A}, \,\cdot\,]$. 
The AST gauge transformation is parameterized by 
a ${\cal G}$-valued vector superfield 
$\Omega(x,\th,\thb)=\Omega^i(x,\th,\thb)T_i$ 
satisfying the constraint ${\Omega^i}\dagg =\Omega^i$, 
\beq
 && \delta B_{\al} = - \2{4} \bar D \bar D 
    {\cal D}_{\al} (e^{-\lam A}\Omega), \hs{10} 
 \delta \bar B_{\dot\al} 
 = - \2{4} D D \bar {\cal D}_{\dot\al} (\Omega e^{-\lam A}),\non
 && \delta A = 0. \label{SUSY-TGT}
\eeq
This transformation is Abelian, 
though $\Omega$ is ${\cal G}$-valued. 
The Lagrangian (\ref{Lag.}) is invariant under 
the {\it global} $G$-transformation
\beq
 &&B_{\al} \to B_{\al}' = g^{-1} B_{\al} g, \hs{10} 
   \bar B_{\dot\al} \to \bar B_{\dot\al}' 
   = g^{-1} \bar B_{\dot\al} g, \non 
 && A \to A' = g^{-1} A g , \hs{10} 
    W_{\al} \to W_{\al}' = g^{-1} W_{\al} g , 
 \label{G-action}
\eeq
with $g \in G$.

The equation of motion for the auxiliary superfield $A$ 
has the complicated form~\cite{LR}  
\beq
  {\cal D}^{\al} B_{\al} 
 + e^{-\lam A} \bar {\cal D}_{\dot\al} 
   \bar B^{\dot\al} e^{\lam A} = - A, \label{EOM_A}
\eeq 
where 
${\cal D}_{\al} B_{\beta} \equiv  D_{\al}B_{\beta} 
+ \{ e^{-\lam A} D_{\al} e^{\lam A}, B_{\beta} \}$. 
If we eliminate $A$ by solving Eq.~(\ref{EOM_A}), 
we obtain the second-order Lagrangian for $B_{\al}$, 
which we call the `supersymmetric theory 
of the FT model'.  
In practice it is difficult to solve Eq.~(\ref{EOM_A}) explicitly, 
and hence we do not write the Lagrangian for $B_{\al}$ explicitly.

On the other hand, if we eliminate $B_{\al}(x,\th,\thb)$,  
we can obtain the dual NLSM as follows.  
The equation of motion for $B_{\al}(x,\th,\thb)$ reads 
\beq
 -4\lam W_{\al}(x,\th,\thb) 
  = \bar D \bar D (e^{-\lam A}D_{\al}e^{\lam A}) 
  = 0, \label{vanish}
\eeq
expressing the fact that $A$ is a pure gauge. 
The solution is written as 
\beq
 e^{\lam  A(x,\th,\thb)} 
 = e^{\phi\dagg(x,\th,\thb)} e^{\phi(x,\th,\thb)}, \hs{10} 
 \bar D_{\dot\al} \phi(x,\th,\thb) = 0 .  \label{pure_gauge}
\eeq
Here $\phi=\phi^i T_i$ is a ${\cal G}$-valued chiral superfield.  
Taking account of Eq.~(\ref{vanish}) and 
substituting (\ref{pure_gauge}) back into the Lagrangian (\ref{Lag.}), 
we obtain the Lagrangian for $\phi$~\cite{CLL}: 
\beq
 {\cal L} = \int d^4\th \; K (\phi,\phi\dagg)
 = \int d^4\th \; \1{4c \lam^2} \tr \log^2 (e^{\phi\dagg} e^{\phi}). 
 \label{nlsm_on_GC}
\eeq
We have thus obtained the NLSM that is dual to 
the supersymmetric theory of the FT model. 

In the Lagrangian (\ref{nlsm_on_GC}), 
$K$ is the K\"ahler potential of the target space of the NLSM.
It can be expanded in powers of $\phi^i$ and ${\phi^i}\dagg$.   
We have 
\beq
 && {\cal L} 
 = \int d^4\th 
  \1{2 \lam^2} \left[\phi^{i\dagger}\phi^i 
 - \1{24} {f_{ml}}^i{f_{jk}}^l 
   (\phi^{m\dagger}\phi^{j\dagger}\phi^k\phi^i \right.\non
&& \left. \hs{35} 
   + 2\phi^{i\dagger}\phi^{m\dagger}\phi^{j\dagger}\phi^k
   + 2\phi^{k\dagger}\phi^i\phi^j\phi^m)
 + \cdots \right] \, .
\eeq
We finds that the dual field theory is the free Wess-Zumino model 
if the group $G$ is Abelian. 

Since the scalar components of $\phi^i$ are complex, 
the target space of this NLSM is $\GC$, 
the complex extension of $G$. 
The Lagrangian (\ref{nlsm_on_GC}) is invariant under 
the global action of the group $G \times G$ :
\beq
 e^{\phi} \to e^{\phi\pri} = g_L e^{\phi} g_R, \hs{10}
 (g_L,g_R) \in G \times G . \label{global}
\eeq
Note that the Lagrangian (\ref{nlsm_on_GC}) is not invariant under 
$\GC\times\GC$, though the target space is $\GC$. 
We find from Eqs.~(\ref{G-action}) and (\ref{pure_gauge}) 
that the right action of Eq.~(\ref{global}) arises from 
the original global symmetry $G$ of 
the AST gauge theory. 
By contrast, 
the left action of Eq.~(\ref{global}) 
is a {\it hidden} global symmetry, 
preserving Eq.~(\ref{pure_gauge}).

The Lagrangian (\ref{nlsm_on_GC}) allows a simple 
interpretation from the viewpoint of 
the supersymmetric nonlinear realization of a global symmetry. 
In supersymmetric field theories, 
spontaneous symmetry breaking is caused 
by the superpotential. 
Since the superpotential is a holomorphic function of 
chiral superfields, its symmetry $G$ is promoted to $\GC$.
Correspondingly, 
there appear quasi-Nambu-Goldstone (QNG) bosons 
in addition to ordinary Nambu-Goldstone (NG) bosons~\cite{KOY}. 
The low-energy effective Lagrangian of these massless bosons 
and their fermionic superpartners  
form a supersymmetric NLSM 
whose target manifold is a K\"ahler manifold~\cite{Zu}.  
This manifold is non-compact due to the existence of QNG bosons~\cite{Sh}. 
Since the target space of the Lagrangian (\ref{nlsm_on_GC}) 
is $\GC$, The NG and QNG bosons are equal in number 
for completely broken $G$. 

We now examine the relation between 
the fields in the AST gauge theory 
and the NG and QNG bosons in the dual NLSM. 
This relation can be worked out explicitly in 
the Abelian case as follows. 
The AST gauge transformation of 
$B_{\mu\nu}$ is given by 
$\delta B_{\mu\nu}= \del_{\mu}\xi_{\nu} - \del_{\nu}\xi_{\mu}$, 
where $\xi_{\mu}$ is the vector component of 
the parameter vector superfield $\Omega$.
Using the AST gauge transformation of the other fields, 
we can take the Wess-Zumino gauge, in which $\psi_{\al}=D=0$. 
The second-order action for $B_{\al}$, 
obtained by eliminating $A$ in the Lagrangian (\ref{Lag.}) 
for the Abelian case, is given by~\cite{Si} 
\beq
 S = - \int d^4x \int d^4\th \, \1{2} G^2 
 = \int d^4x \left(- \1{2}\del_{\mu}C \del^{\mu}C 
  - i\lamb \sigb^{\mu}\del_{\mu}\lam 
  + \1{4}G^{\mu}G_{\mu} \right).
\eeq
Here $G(x,\th,\thb)$ is the field strength of 
$B_{\al}$ defined by 
$G \equiv \1{2}(D^{\al}B_{\al}+\bar D_{\dot\al}\bar B^{\dot\al})
= \th\sig^{\mu}\thb G_{\mu} + \cdots$, 
where $G^{\mu} \equiv \eps^{\mu\nu\rho\sig} 
\del_{\nu}B_{\rho\sig}$.
We find from this action that 
$B_{\mu\nu}$ corresponds to the NG boson  
and $C$ to the QNG boson. 
For the non-Abelian case, the second-order Lagrangian 
is an interacting field theory, 
and the relation between NG bosons and 
$B_{\mu\nu}$ and that between QNG bosons and 
$C$ are less straightforward.

\section{Supersymmetric extension of the non-Abelian 
scalar-tensor duality for coset spaces}

\subsection{Anti-symmetric tensor gauge theory coupled to 
Yang-Mills fields}
We now generalize the scalar-tensor duality 
to the case of coset space. 
In supersymmetric theories, 
we introduce the YM vector superfield  
$V(x,\th,\thb)= V^a(x,\th,\thb) H_a$,  
where the $H_a$ are generators of a {\it subgroup} $H$ of $G$. 
We decompose the Lie algebra as in Eq.~(\ref{generator}).

As mentioned above, there are 
two expressions we can use to generalize 
the FT model to the case of coset space. 
We use the second expression (\ref{coset1''}).
In this case, we do not need to modify 
the AST gauge transformation and 
the field strength. 

In analogy to the YM gauge transformation of $A_{\mu}$ in the bosonic case, 
the transformation of $A$ and the YM gauge field $V$ are closely related, with 
$\lam A$ and $e V$ transforming identically.
The YM gauge transformation is
\beq
 && B_{\al} \to  B_{\al}' = e^{-i \Lam} B_{\al} e^{i \Lam}, \hs{14} 
    \bar B_{\dot\al} \to \bar B_{\dot\al}' 
   = e^{-i \Lam\dagg} \bar B_{\dot\al} e^{i \Lam\dagg}, \non
 && e^{e V} \to e^{e V'} 
  = e^{-i \Lam\dagg} e^{e V} e^{i \Lam}, \hs{10}
    e^{\lam A} \to e^{\lam A'} 
  = e^{-i \Lam\dagg} e^{\lam A} e^{i \Lam}, 
  \label{gauge_tr.}
\eeq
where the gauge parameter is an ${\cal H}$-valued 
chiral superfield $\Lam(x,\th,\thb) = \Lam^a(x,\th,\thb) H_a$, 
satisfying $\bar D_{\dot\al} \Lam (x,\th,\thb)= 0$. 
The field strengths of $A$ are 
the same as those in (\ref{Auxiliary-fs}). 
They transform as
\beq
  W_{\al} \to W_{\al}' = e^{-i \Lam} W_{\al} e^{i \Lam},\hs{10}    
    \bar W_{\dot\al} \to \bar W_{\dot\al}' 
     = e^{-i \Lam\dagg} \bar W_{\dot\al} e^{i \Lam\dagg},
 \label{gauge_tr.2}
\eeq
under the YM gauge transformation (\ref{gauge_tr.}). 

In Eq.~(\ref{Lag.}) we have already given the Lagrangian in first-order form 
in the case in which there is no YM gauge field.  
A generalization of the Lagrangian to that 
in which $G$ is fully gauged by 
supersymmetric YM field, 
is given in Ref.~\cite{CLL}. 
The model investigated there contains the kinetic term of the YM field.
The dual field theory is a massive 
YM field theory.  

We construct the model in which the subgroup 
$H$ of $G$ is gauged by the YM field. 
As in the bosonic case, the model we consider here does not contain 
the kinetic term of the YM field.
The Lagrangian for this model is 
\beq
 {\cal L} 
 &=& - \1{2c} \left[\int d^2\th \; \tr (W^{\al} B_{\al}) 
    + \int d^2 \thb \; 
      \tr (\bar W_{\dot\al} \bar B^{\dot\al}) \right]\non
 && + \1{4 c\lam^2} \int d^4\th\; 
   \tr \log^2 (e^{- e V} e^{\lam A}) . \hs{2}
  \label{gauged-1st-order}
\eeq
If we set $V=0$, we recover the Lagrangian~(\ref{Lag.}). 
We should note that the global action of $G$, 
Eq.~(\ref{G-action}), is explicitly {\it broken} 
by the partial gauging of the subgroup $H$ of $G$; 
the ${\cal H}$-valuedness of $V$ is not maintained under 
the $G$-action $V \to V' = g^{-1} V g$.  

The equation of motion of the auxiliary field $A$ 
is difficult to solve, and we do not write 
the second-order Lagrangian for $B_{\al}$. 

\medskip
Here we discuss the field redefinition 
corresponding to (\ref{redef}) in the bosonic case. 
Using the second alternative, 
corresponding to (\ref{coset1''}), 
the AST gauge transformation 
and the field strengths are not modified.
The Lagrangian for the first alternative, 
corresponding to (\ref{coset1'}), 
is obtained from the Lagrangian (\ref{gauged-1st-order}) 
by the field redefinition of $A$
\beq
 e^{\lam A} \to e^{e V} e^{\lam A}. \label{redef-SUSY}
\eeq 
Using the Hausdorff formula, this can be rewritten as
\beq
 A \to A + {e \over \lam} V + {e \over 2} [A,V] + \cdots.
\eeq
We obtain the Lagrangian as 
\beq
 {\cal L} 
 = - \1{2c} \left[\int d^2\th \; \tr (\tilde W^{\al} B_{\al}) 
    + \int d^2 \thb \; 
      \tr (\tilde {\bar W}_{\dot\al} \bar B^{\dot\al}) \right]
  + \1{4c} \int d^4\th\; \tr A^2 .\hs{2}
  \label{gauged-1st-order2}
\eeq
Here $\tilde W^{\al}$ and $\tilde{\bar W}_{\dot\al}$ are defined by 
\beq
 && \tilde W_{\al} = - \1{4 \lam} \bar D \bar D 
  \left[ e^{-\lam A} e^{-eV} D_{\al}(e^{eV} e^{\lam A}) \right] ,\non
 && \tilde{\bar W}_{\dot\al} 
  = \1{4 \lam} D D 
  \left[ e^{eV} e^{\lam A} \bar D_{\dot\al}(e^{-\lam A}e^{-eV})\right], 
 \label{Auxiliary-fs2}
\eeq 
which are obtained from (\ref{Auxiliary-fs}) 
using (\ref{redef-SUSY}). 
The AST gauge transformation (\ref{SUSY-TGT}) 
is also modified by (\ref{redef-SUSY}) to 
\beq
 && \delta B_{\al} = - \2{4} \bar D \bar D 
    \tilde {\cal D}_{\al} (e^{-\lam A} e^{-eV}\Omega), \hs{10} 
 \delta \bar B_{\dot\al} 
 = - \2{4} D D \tilde{\bar {\cal D}}_{\dot\al} 
    (\Omega e^{-\lam A}e^{-eV}),\non
 && \delta A = 0, \hs{10} \delta V = 0, 
\eeq
with $\tilde {\cal D}_{\al} = D_{\al} 
+ [e^{-\lam A}e^{-eV} D_{\al}(e^{eV} e^{\lam A}), \,\cdot\,]$. 
The YM gauge transformation of $A$ becomes  
\beq
 A \to A' = e^{-i\Lam} A e^{i\Lam}. 
\eeq 
At first sight, it might appear that this is inconsistent 
with the reality condition, 
but this is not the case, as 
$e^{-eV}e^{\lam A}$ and $V$ satisfy the reality condition    
under the redefinition (\ref{redef-SUSY}).

\subsection{Dual nonlinear sigma model on complex coset spaces}
To obtain the dual NLSM, 
we return to the Lagrangian (\ref{gauged-1st-order}). 
Elimination of $B_{\al}(x,\th,\thb)$ 
through its equation of motion again gives 
\beq
 e^{\lam A(x,\th,\thb)} 
 = e^{\phi\dagg(x,\th,\thb)} e^{\phi(x,\th,\thb)}, \hs{10} 
 \bar D_{\dot\al} \phi(x,\th,\thb) = 0 , 
\eeq
where $\phi$ is a ${\cal G}$-valued chiral superfield. 
Proceeding in the same way as in \S3, we obtain the NLSM 
whose K\"ahler potential is
\beq
 K(\phi,\phi\dagg,V) 
 = \1{4c \lam^2} \tr \log^2 (e^{-e V} e^{\phi\dagg} e^{\phi}).
 \label{G/H} 
\eeq

\medskip
As in the bosonic case, the NLSM (\ref{G/H}) is invariant 
under the global transformation of $G$
\beq
 e^{\phi} \to e^{\phi\pri} 
 = g e^{\phi} , \hs{10} g \in G \label{global_ac.}
\eeq
and the gauge transformation of $H$
\beq
 && e^{e V} \to e^{e V'} = h\dagg e^{e V} h, \hs{10} 
    e^{\phi} \to e^{\phi'} = e^{\phi} h, \non
 && h = e^{i\Lam^a(x,\th,\thb)H_a} , \hs{10} 
    \bar D_{\dot\al} \Lam^a(x,\th,\thb) = 0.  \label{local_ac.}
\eeq 
The global action (\ref{global_ac.}) of $G$ corresponds to 
the left action in Eq.~(\ref{global}), 
which does not correspond to any symmetry in the AST gauge theory. 
The local action (\ref{local_ac.}) of $H$ 
corresponds to the right action in Eq.~(\ref{global}), 
and its origin is, of course, 
the gauge transformation (\ref{gauge_tr.}). 
Equation (\ref{G/H}) can be interpreted as the Lagrangian of 
the dual NLSM 
formulated in terms of the {\it hidden local symmetry}~\cite{BKY}. 

The physical degrees of freedom can be found as follows.  
First, decompose $\phi(x,\th,\thb)$ into 
its ${\cal H}$-valued part and the rest as  
\beq
 && e^{\phi(x,\th,\thb)} = \xi(x,\th,\thb) h(x,\th,\thb), \non
 && \xi = e^{i \ph^I(x,\th,\thb) X_I},  \hs{10} 
    h = e^{i\al^a (x,\th,\thb) H_a}, \label{dec.}
\eeq
where the $X_I$ are the coset generators 
and $\ph^I(x,\th,\thb)$ and $\al^a(x,\th,\thb)$ 
are chiral superfields. 
By fixing the gauge as
\beq
 e^{e V} = h\dagg e^{e V_0} h, \hs{10} 
 h = e^{i\al^a (x,\th,\thb) H_a}, 
\eeq
we obtain the K\"ahler potential, 
\beq
 K(\phi,\phi\dagg,V_0) 
 = \1{4c \lam^2} \tr \log^2 (e^{-e V_0} \xi\dagg\xi). \label{G/H2}
\eeq
This Lagrangian is invariant under 
the global $G$ transformation, defined by 
\beq
 && \xi \to \xi' = g \xi h(g,\xi) , \non
 && e^{e V_0} \to e^{e {V_0}'} 
  = h\dagg(g,\xi) e^{e V_0} h(g,\xi), \hs{10} g \in G, \label{CV}
\eeq
where the $H$-valued chiral superfield 
$h(g,\xi(x,\th,\thb))$ is a compensator needed to 
recover the decomposition in Eq.~(\ref{dec.}). 

The superfield $\xi$ in (\ref{CV}) is a coset representative of $\GC/\HC$ and 
the bosonic parts of $\ph^I(x,\th,\thb)$ 
parameterize $\GC/\HC$. 
The superfield $\al^a(x,\th,\thb)$ in Eq.~(\ref{dec.}) 
has been absorbed into $V^a$ 
as a result of the supersymmetric Higgs mechanism. 
Each chiral superfield $\ph^I$ in this NLSM 
contains the same number of 
NG and QNG bosons. 
The relation to the general theory of 
supersymmetric nonlinear realizations~\cite{BKMU} is, 
however, obscure at this stage,  
since we are unable to eliminate $V$ explicitly.

\section{Discussion}

We have constructed a supersymmetric extension of 
the duality between AST gauge theories and the NLSM on coset spaces. 
The two models have apparently 
different manifest symmetries.

It is useful to clarify the relations between 
the symmetries of the two models. 
We do this explicitly in the bosonic case and simply note that 
the situation is the same in the supersymmetric case. 
The relations between symmetries of the AST gauge theories and 
the NLSM are listed in Table 1 for both the FT model 
and the $H_R$-gauged FT model.

\begin{footnotesize}
\begin{table}[htbp]
\begin{center}
\caption{\footnotesize 
Relations between the symmetries of the AST gauge theories 
and the NLSM. The 
upper table corresponds to the FT scalar-tensor duality. 
The $G/H$ scalar-tensor duality, corresponds to the lower table, is obtained 
by gauging the FT model. 
\label{Fig1}}
\vspace{5mm}
\begin{tabular}{|c|ccccc|}
\hline
model& FT model & $\longleftrightarrow$ & NLSM on $G$ & &\\
\hline
symmetry& {\it global} $G_R \times${\it AST gauge} & 
        & {\it global} $G_L \times${\it global} $G_R$ & &\\ 
\hline
\hline
model& $H_R$-gauged FT model & $\longleftrightarrow$ &$H_R$-gauged 
NLSM on $G$ & $\longleftrightarrow$ & $G/H$ NLSM\\
\hline
symmetry& {\it local} $H_R \times${\it AST gauge} & 
        &{\it global} $G_L \times${\it local} 
$H_R$ & & {\it global} $G_L$\\ 
\hline
\end{tabular}
\end{center}
\end{table}
\end{footnotesize}

\noindent This table deserves several comments:

\noindent i) The AST gauge symmetry is hidden in the gauged NLSM.

\noindent ii) The global symmetry $G_R$, 
Eq.~(\ref{Gglobe}) [(\ref{G-action}) in the SUSY case], of 
the FT model is broken in the gauged FT model.

\noindent iii) The global symmetry $G_L$, 
Eq.~(\ref{globalG}) [(\ref{global_ac.}) in the SUSY case] 
of the gauged NLSM is hidden in the gauged FT model.

\noindent iv) 
The local symmetry $H_R$, Eq.~(\ref{localH}) [(\ref{local_ac.}) in the SUSY case],
of the gauged NLSM is hidden in the NLSM on $G/H$.

\noindent i) and iii) are also true in the non-gauged models.

In this paper, we consider the specific form (\ref{Lag.}) 
of the supersymmetric Lagrangian. The last term of (\ref{Lag.}) 
can be generalized to 
$$\int d^4 \th \; F(A)=\int d^4 \th \; f(\tr e^{\lam A},\ \tr e^{2\lam A},\ \tr e^{3\lam A},\ \cdots)\ ,$$ where
$F (A)$ is a general $G$-invariant function of $A$~\cite{BT}. 
In this case, we obtain the NLSM with the general form of the 
K\"ahler potential 
$$K (\phi,\phi\dagg) = f (\tr (e^{\phi\dagg} e^{\phi}),\ 
\tr (e^{\phi\dagg} e^{\phi})^2,\ \tr (e^{\phi\dagg} e^{\phi})^3,\ \cdots),$$ 
instead of (\ref{nlsm_on_GC}). The arbitrariness of the K\"ahler
potential reflects the fact that the metric of the target 
space $\GC$ is not completely determined by the symmetry 
$G \times G$, and it is related to the existence of QNG bosons.

The existence of solitons plays a role in the duality 
in supersymmetric gauge theories~\cite{SUSY-QCD}. 
This suggests that perhaps solitons also play a role 
in the scalar-tensor duality. From this point of view 
it is of interest to extend the scalar-tensor duality 
by including higher-order derivative terms in 
the NLSM and the supersymmetric NLSM~\cite{Ni}.

\section*{Acknowledgements}
We would like to thank Kiyoshi Higashijima and Akio Sugamoto 
for valuable comments. 
K. F. and H. N. are supported by a Research 
Assistant Fellowship of Chuo University. This work is supported 
partially by grants of Ministry of Education, 
Science and Technology, (Priority Area B, ``Supersymmetry and 
Unified Theory" and Basic Research C). 


\end{document}